\documentstyle[12pt]{article}
\newcommand{\beq}{\begin{equation}}
\newcommand{\eeq}{\end{equation}}
\newcommand{\beqa}{\begin{eqnarray}}
\newcommand{\eeqa}{\end{eqnarray}}
\newcommand{\beqan}{\begin{eqnarray*}}
\newcommand{\eeqan}{\end{eqnarray*}}
\newcommand{\no}{\nonumber}

\newcommand{\ul}{\underline}
\newcommand{\ol}{\overline}
\newcommand{\ra}{\rightarrow}
\newcommand{\lra}{\longrightarrow}

\newcommand{\ben}{\begin{enumerate}}
\newcommand{\een}{\end{enumerate}}
\newcommand{\bfl}{\begin{flushleft}}
\newcommand{\efl}{\end{flushleft}}
\newcommand{\ba}{\begin{array}}
\newcommand{\ea}{\end{array}}
\newcommand{\btab}{\begin{tabular}}
\newcommand{\etab}{\end{tabular}}
\newcommand{\bit}{\begin{itemize}}
\newcommand{\eit}{\end{itemize}}

\newcommand{\vs}{\vspace}
\newcommand{\hs}{\hspace}
\newcommand{\prepr}[1] {\begin{flushright} {\bf #1} \end{flushright} \vskip
1.5cm}
\newcommand{\titul}[1] {\begin{center}{\large \bf #1 } \end{center}\vskip 1.cm}

\newcommand{\autor}[1] {\begin{center} {\bf \lineskip .3cm #1  }
                        \end{center} }

\newcommand{\lugar}[1] {\begin{center}  {\large \it #1   } \end{center}}
\newcommand{\abstr}[1] {{\begin{center} \vskip .5cm {\bf Abstract
                        \vspace{0pt}} \end{center}}\begin{quote} #1
                        \end{quote}}
\newcounter{muni}

\begin{document}
\vspace{.4cm}
\hbadness=10000
\pagenumbering{arabic}
\begin{titlepage}
\prepr{Preprint hep-ph/9405XXX \\PAR/LPTHE/94-22  \\26 May 1994 }
\titul{ ON THE DETERMINATION OF $a_1$ AND $a_2$ FROM
 HADRONIC TWO BODY $B$ DECAYS.}
\autor{ M. Gourdin\footnote{\rm Postal address: LPTHE, Tour 16, $1^{er}$ Etage,
Universit\'e Pierre {\it \&} Marie Curie and Universit\'e Paris VII, 4 Place
Jussieu, F-75252 Paris CEDEX 05, France.},
\hs{3mm} A. N. Kamal$^{1,}\footnote{Permanent address : Department of Physics,
University of Alberta, Edmonton, Alberta T6G2JI, Canada }$ }
\autor{ Y. Y. Keum$^1$ and X. Y. Pham$^1$ }

\lugar{Laboratoire de Physique Th\'eorique et Hautes Energies,\footnote{\em
Unit\'e associ\'ee au CNRS URA 280}
 Paris, France }

\begin{center}
{\small  E-mail : gourdin@lpthe.jussieu.fr, keum@lpthe.jussieu.fr,
pham@lpthe.jussieu.fr, \\ kamal@lpthe.jussieu.fr and
kamal@bose.phys.ualberta.ca}
\end{center}

\vs{-12cm}
\thispagestyle{empty}
\vs{120mm}
\noindent
\abstr{
{}From Class I decays : $B^o \ra \pi^+ ( \rho^+ ) + D^- ( {D^*}^- )$, we
determine $a_1$, and from Class III decays : $B^+ \ra \pi^+ ( \rho^+ ) +
\ol{D}^o ( {\ol{D}^*}^o )$,
we determine an allowed domain in the $( a_1, a_2 )$ plane. We find that within
one standard deviation errors, the allowed band of $a_1$ from Class I decays is
at least three standard deviations removed from the allowed domain $( a_1, a_2
)$ from Class III decays.If we expand the experimental errors to two standard
deviations we do find a small intersection between the $a_1$ band and the
allowed $( a_1, a_2 )$ domain.
The results usually quoted in the literature lie in this intersection.
We suggest :
(1) an independent measurement of the branching ratio for the Class III decay,
$B^+ \ra \rho^+ \ol{D}^o $,
(2) a high-statistics measurement of the branching ratio of the Class I decay,
$B \ra \ol{D} ( \ol{D}^* ) +  D_s ( D_s^* )$ in both charged states, and
(3) a measurement of the longitudinal polarization fraction in the Class III
decay
$B^+ \ra \rho^+ {\ol{D}^*}^o $ to shed more light on the questions involved .
}
%
\end{titlepage}

\newpage
 In recent past there has been considerable interest [1-4] in the values, and
the relative signs, of the phenomenological parameters $a_1$ and $a_2$
introduced by Bauer, Stech and Wirbel (BSW henceforth) \cite{BSW} in $B$
decays.
The coefficient $a_1$ is determined from  $ B^o \ra \pi^+(\rho^+) + D^-
({D^*}^-) $ modes which are  Class I \cite{BSW} processes.

Decay amplitudes for $ B^+ \ra \pi^+(\rho^+) + \ol{D}^o ({\ol{D}^*}^o) $ which
are of Class III \cite{BSW}  depend on both $a_1$ and $a_2$.
Thus, independent of the above Class I decays, those of Class III provide a
determination of an allowed region in $( a_1, a_2 )$ plane.

In principle, $ B^o \ra \pi^o (\rho^o) + \ol{D}^o ({\ol{D}^*}^o) $, which are
Class II
\cite{BSW} decays, could determine $|a_2|$ independently also.
However, only upper limits exist on the branching ratios of these
colour-suppressed modes \cite{Browder}.
Instead, $|a_2|$ is generally determined from the only colour-suppressed $B$
decays
that have been observed, namely those involving charmonium states such as $B
\ra J/\Psi K$ and $J/\Psi K^*$.
We discuss the extraction of $|a_2|$ from these decays also, albeit, we suggest
caution in using these modes for such a determination as explained later.

We begin by describing our calculational procedure and the parameters used.
We assume  factorization \cite{BSW}.
We do not discuss the effect of final state interactions (FSI) in this paper.
On physical grounds FSI effects are expected to be small as the produced
quarks, being extremely relativistic, leave the strong interaction region
before hadronization so that FSI
between hadrons do not play a significant role.
Ref. \cite{KP} provides an estimate of the strong interaction phase angles
which turn out to be small.

The parameters we use are \cite{Browder} :
\beqa
& & \tau_{B^+} = \tau_{B^o} = \tau_{B} = 1.44 \hs{2mm}\cdot 10^{-12} s,
\hs{20mm}
|V_{cb}| = 0.041  \no \\
& & \label{eq1} \\
& & f_{\pi^+} = 131.7 MeV, \hs{8mm}f_{\rho^+} = 212 MeV, \hs{8mm}f_{D} =
f_{D^*} = 220 MeV. \no
\eeqa

Note that if we had used $\tau_B = 1.18 \cdot 10^{-12} s $ and $|V_{cb}| =
0.045$ as in \cite{Neubert}, the coefficients $a_1$ and $a_2$ would be scaled
down by $0.65 \%$.

The other theoretical inputs in our calculation are the hadronic form factors
for $B \ra D (D^*)$ and $B \ra \pi (\rho)$ transitions for which we use various
models as will be described in the text.

As for the experimental input, throughout this paper we use the World average
value of data quoted in \cite{Browder}.
This average is dominated by the new high-statistics CLEO II data
\cite{Browder}.

\vs{4mm}
We start our analysis with the four Class I decays
\beq
 B_d^o \hs{1mm} \lra \hs{1mm} \pi^{+} (\rho^{+}) \hs{2mm} + \hs{2mm} D^{-}
({D^{*}}^{-})  \label{eq2}
\eeq
The Cabibbo-Kobayashi-Maskawa (CKM) factor is $V_{cb}^{*}V_{ud}$ and the decay
amplitude is the sum of two contributions from the spectator and W-exchange
diagrams.
To our knowledge the W-exchange diagram has never been computed in a
satisfactory way and, on the basis of reasonable arguments,  it has been
neglected in $B$ meson decay analysis.
We disregard it in the present analysis, however, we shall come back to this
point later.

The spectator amplitude involves the $B \ra D$ and $B \ra D^*$ hadronic form
factors which correspond to the heavy quark to heavy quark decay.
We shall consider the following three sets of form factors : \vs{2mm}

\hs{10mm} (a) The set BSW as in the original Bauer-Stech-Wirbel model
\cite{BSW}. \vs{2mm}

\hs{10mm} (b) The set HQET I associated with exact heavy quark symmetry and
used by Deandrea et al \cite{Gatto} in their analysis.
An extrapolation of the Isgur-Wise function $\xi(y)$ is made from the symmetry
point using  an improved form of the relativistic oscillator model as described
in \cite{NR}. \vs{2mm}

\hs{10mm} (c) The set HQET II associated with heavy quark symmetry including
mass corrections as proposed by Neubert et al \cite{Neubert}. We have made an
interpolation of the entries in Table 4 of \cite{Neubert} to obtain the form
factors needed.

The  branching ratios for these Class I decays can be  written in the form
\beq
{\cal B}( B_d^o \ra f^o ) = C_f^o \hs{2mm} a_1^2 \hs{2mm} 10^{-2} \label{eq3}
\eeq
where the coefficients $C_f^o$ are given in  Table 1.

\vs{5mm}
\begin{center}
\begin{tabular}{|c||c|c|c|} \hline
f & BSWI \cite{BSW}   & HQET I \cite{Gatto} &  HQET II \cite{Neubert} \\
\hline\hline
$\pi^+ D^-$ & 0.384 & 0.271 & 0.271 \\
\hline
$\rho^+ D^-$ & 0.931 & 0.674 & 0.683 \\
\hline
$\pi^+ {D^*}^-$ & 0.292 & 0.288 & 0.262 \\
\hline
$\rho^+ {D^*}^-$ & 0.879 & 0.844 & 0.784 \\
\hline
\end{tabular} \\

\vs{5mm}
$\ul{\rm Table \hs{2mm} 1.}$ \vs{2mm}

{\small Coefficients $C_f^o$ for $B_d^o \hs{1mm} \lra \hs{1mm} \pi^{+}
(\rho^{+}) \hs{2mm} + \hs{2mm} D^{-} ({D^{*}}^{-}) $ }

\end{center}
 Using the World average data given by Browder et al \cite{Browder}, we obtain
the following limits from the intersection of the one standard deviation ranges
for the four decays,
\beqa
& & 0.91 \leq a_1 \leq 0.97 \hs{25mm} {\rm BSW} \no \\
\cr
& & 0.96 \leq a_1 \leq 1.03 \hs{25mm} {\rm HQET \hs{2mm} I} \label{eq4} \\
\cr
& & 0.96 \leq a_1 \leq 1.07 \hs{25mm} {\rm HQET \hs{2mm} II \hs{2mm}( one
\hs{2mm} standard \hs{2mm} deviation )} \no \\
\cr
& & 0.87 \leq a_1 \leq 1.17 \hs{25mm} {\rm HQET \hs{2mm} II \hs{2mm}( two
\hs{2mm} standard \hs{2mm} deviations )} \no
\eeqa

The value of $a_1$ clearly does not depend sensitively on the choice of the
model for $B \ra D (D^*)$ transition form factors, and with the neglect of
W-exchange contributions, $a_1$ is not expected to differ from unity by more
than $10 \%$.

We observe that the coefficients $C^o_f$ of the Table 1 differ slightly from
those given in previous analyses \cite{{Neubert},{Browder},{Gatto}} because of
different choice for $\tau_B$, $|V_{cb}|$ and $f_{\rho}$.

We finally add a comment on recent CLEO II data on the longitudinal
polarization in the Class I decay,
$B_d^o \ra \rho^+ + {D^*}^-$.
The relative amount of longitudinal polarization, $\Gamma_L/ \Gamma $, has been
measured to be \cite{{Browder},{Besson}}
\beq
\frac{\Gamma_L} {\Gamma }|_{exp} = 0.90 \pm 0.07 \pm 0.05 \label{eq5}
\eeq
which is in perfect agreement with the theoretical predictions of all the three
models ; 0.87 for BSW, 0.88 for HQET I and HQET II.

\vs{4mm}
We now turn to a study of the following Class III decays,
\beq
 B^{+}_u \lra \pi^{+} (\rho^+) + \ol{D}^o ( \ol{D}^{{o}^*} )  \label{eq6}
\eeq
Again the CKM factor is $V_{cb}^*V_{ud}$ and the decay amplitude is the sum of
two contributions, one  due to the spectator diagram and the other due to the
colour-suppressed diagram.
The spectator amplitude is the same as for Class I decay but for minor
corrections due to the $\ol{D}^o - D^{-}$
or ${\ol{D}^o}^* - {D^{-}}^*$ mass difference.

The colour-suppressed amplitude involves $B \ra \pi$ and $B \ra \rho$ hadronic
form factors which correspond to the heavy quark to light quark transition,  to
which there is no guidance from  heavy quark symmetry. We shall consider here
three sets of form factors  similar to those used  in a recent work
\cite{GKP}.

\hs{10mm} (a) The set BSW I which is the original Bauer-Stech-Wirbel model
\cite{BSW} where all $q^2$ dependence are of the monopole type.

\vs{2mm}
\hs{10mm} (b) The set BSW II which is a slightly modified version of BSW I used
in \cite{Neubert}.
The normalization at $q^2 = 0 $ and the pole masses are unchanged, however, the
form factors $F_0$ and $A_1$  still
have a monopole $q^2$-dependence but the form factors $F_1, A_0, A_2 $ and $V$
have a dipole $q^2$-dependence.

\vs{2mm}
\hs{10mm} (c) The set CDDFGN of Ref. \cite{CDD} used in the analysis of
Deandrea et al \cite{Gatto}, where the normalization of the heavy to light
transition form factors at $q^2 = 0$ has been estimated in a model combining
chiral and heavy quark symmetry with mass corrections.
The pole masses are as in \cite{BSW} and a monopole $q^2$-dependence is used.

The branching ratios for these Class III decays are written in the form,
\beq
{\cal B}( B_u^+ \ra f ) = C^+_{f} \hs{1.5mm} a_1^2 \hs{1.5mm}( 1 +
\frac{a_2}{a_1} \ I_f )^2 \cdot 10^{-2}\label{eq7}
\eeq
for $f = \pi^+ \ol{D}^o, \rho^+ \ol{D}^o $ and $\pi^+ {\ol{D}^*}^o$, and for $f
= \rho^+ {\ol{D}^*}^o$,
\beq
{\cal B}( B_u^+ \ra \rho^+ {\ol{D}^*}^o ) = C^+_{\rho^+{\ol{D}^*}^o} \hs{1.5mm}
a_1^2 \hs{1.5mm} [1 + 2 \frac{a_2}{a_1} \ I_{\rho^+{\ol{D}^*}^o} +
\frac{a_2^2}{a_1^2} \ J^2
]\ \cdot 10^{-2}  \label{eq8}
\eeq

Of course, the coefficients $C^+_{f}$ for $B^+$ decay and $C^o_{f}$ for $B^o$
decay, both arising from the spectator amplitude, differ only by small phase
space corrections due to the
$\ol{D}^o - D^-$ and ${\ol{D}^*}^o - {D^*}^-$ mass differences which are
typically $O(10^{-3})$.
This difference will be neglected.

The coefficients $I_f$ and $J$ are related to the ratios of colour-suppressed
to spectator amplitudes.
They also involve the leptonic constants $f_D$ and $f_D^*$ for which we use the
estimates in (1). For $B \ra D (D^*)$ transitions we have used the three sets
of form factors, BSW, HQET I and HQET II, as in Class I
decays discussed earlier and for $B \ra \pi (\rho)$ transitions those from BSW
I, BSW II and CDDFGN models.
In Table 2  we show the results for only the case where we have employed HQET
II for the $B \ra D ( D^* )$ hadronic form factors.

\vs{5mm}
\begin{center}
\begin{tabular}{|c||c|c|c|} \hline
f & BSW I    &  BSW II &  CDDFGN \\
\hline\hline
$\pi^+ \ol{D}^o$ & 1.222 & 1.222 & 1.945$^{(a)}$   \\
\hline
$\rho^+ \ol{D}^o$ & 0.556 & 0.635 & 0.475 \\
\hline
$\pi^+ {\ol{D}^*}^0$ & 1.100 & 1.282 & 1.749 \\
\hline
 & 0.865 & 0.708 & 0.740 \\
$\rho^+ {\ol{D}^*}^o$ &  &   & \\
  & J = 0.885 & J = 0.760 &  J = 0.848 \\
\hline
\end{tabular} \\
$(a)$ Our value differs significantly from the value of 1.155 quoted in
\cite{Browder}.

\vs{5mm}
$\ul{\rm Table \hs{2mm} 2.}$ \vs{2mm}

{\small Coefficients $I_f$ and $J$ for $B_u^+ \hs{1mm} \lra \hs{1mm} \pi^{+}
(\rho^{+}) \hs{2mm} + \hs{2mm} \ol{D}^{o} ({\ol{D}^{*}}^{o}) $ }

\end{center}

With Eqs. (7), (8) and the World average data \cite{Browder} we determined the
allowed region in $( a_1, a_2 )$ plane corresponding to one standard deviation
error for all the four decays and found a non-empty intersection of allowed
domain.
We display the result of our analysis of the cases shown in Table 2
in Figure 1 for BSW I, Figure 2 for BSW II and Figure 3 for CDDFGN models.

An interesting qualitative feature of the allowed domain in the $(a_1,a_2)$
plane is that both types of solutions $a_2 > 0$ and $a_2 < 0$ exist.
When $a_2$ is positive $a_1$ is in the 1.2 to 1.4 range, and when  $a_2$ is
negative $a_1$ is in the 1.3 to 1.7 range.
Essentially on this basis it has been claimed that the new CLEO II data favour
an $a_2/a_1$ positive solution \cite{{Neubert},{Stone},{KP},{Browder},{Gatto}}.

If we restrict ourselves to one standard deviation limits and use HQET II form
factors for $B \ra D (D^*)$ transitions then it is evident that the value of
$a_1$ determined from the Class I processes
$B^o \ra \pi^+ (\rho^+) + D^-({D^*}^-)$, $a_1 = 1.01 \pm 0.06$, given in (4),
does not have an overlap with the allowed domain in $( a_1, a_2 )$ plane
determined from the Class III processes, $B^+ \ra \pi^+ (\rho^+) +
\ol{D}^o({\ol{D}^*}^o)$ shown in Figs. 1, 2 and 3.
In fact, they are at least $\ul{\rm three \hs{2mm} standard \hs{2mm} deviations
\hs{2mm} below}$ the $( a_1, a_2 )$ domain allowed by the Class III decays.
There are essentially two ways to improve the situation :

\hs{10mm}(1) Considering, as an example, the point $a_1 = 1.05$, $a_2 =0.25$ we
see from Figs. 1, 2 and 3 that all experiments but $B^+ \ra \rho^+ \ol{D}^o$
can accomodate such a solution.
The theoretical expectation for ${\cal B}(B^+ \ra \rho^+\ol{D}^o)$ with such
value of $a_1$ and $a_2$ is $ (0.97 \pm 0.03) 10^{-2}$ e.g. two standard
deviations below the experimental result from CLEO II.
Therefore, new measurement of $B^+ \ra \rho^+\ol{D}^o$ would be welcome.

\hs{10mm}(2) On the theoretical side, the value of $a_1$ extracted from $B^o
\ra \pi^+(\rho^+) + D^-({D^*}^-)$ experiments can increase if we have an
important W-exchange contribution such that
\beqa
& & Re \hs{1mm} F^{ANN} > 0 \hs{20mm} {\rm if} \hs{10mm} a_2/a_1 < 0  \no \\
& & \label{eq9} \\
& & Re \hs{1mm} F^{ANN} < 0 \hs{20mm} {\rm if} \hs{10mm} a_2/a_1 > 0  \no
\eeqa

We are not aware of a reliable calculation of the complex annihilation form
factors at the large $q^2$ value, $q^2 = m^2_B$, needed here.
{}From the knowledge of the singularity structure of $D \ra \pi (\rho) $ form
factors, it would be naive to expect a monopole extrapolation from the low
$q^2$ region - where they can be measured  in semileptonic $D$ decays - to $q^2
= m^2_B$ to be reliable.
We are presently looking at realistic approaches to this problem.

In order to seek conditions under which solutions for $a_1$ and $a_2$ generally
quoted in the literature \cite{Browder} are obtained, we repeated our analysis
but now with two standard deviation limits.
We show the result in Figure 4 for the case where HQET II is used for $B \ra D
(D^*)$ form factors and BSW II for $B \ra \pi (\rho)$ form factors.
Now we do find a common intersection of the band of allowed values of $a_1$
arising from Class I decays and the allowed $( a_1, a_2 )$ domain from Class
III decays.
This intersection is shown as the shaded wedge in Figure 4.
The solution quoted in \cite{Browder} lies in this wedge.
The boundaries of this wedge allow,
\beq
0.90 \leq a_1 \leq 1.17, \hs{10mm} 0.05 \leq a_2 \leq 0.48, \hs{10mm} 0.04 \leq
\frac{a_2}{a_1} \ \leq 0.54
\label{eq10}
\eeq
However the values of $a_1$ and $a_2$ are strongly correlated;
a large value of $a_2$ demands a smaller value of $a_1$ and vice versa.
Furthermore, as this common intersection does not exist at the one standard
deviation level,
the quality of any attempted simultaneous fit to  extract $a_1$ and $a_2$ from
Class I and III data must be quite poor.

We have at our disposal other Cabibbo-favoured modes to determine the parameter
$a_1$ independently.
They are the Class I decays to which W-exchange cannot contribute,
\beq
B^o_d \lra D^- ( {D^*}^- ) + D^+_s ( {D^*}_s^+ ) \label{eq11}
\eeq
These modes are governed, at the spectator level, by the CKM factor
$V^*_{cb}V_{cs}$.
Of course, penguin contributions are present but we have some control over
their size.
Using HQET II set for $B \ra D ( D^* )$ form factors \cite{Neubert} and  for
the leptonic decay constants a value $f_{D_s} = f_{D^*_s} = 280 MeV $
compatible with the recent experimental data on $D^+_s \ra \mu^+ + \nu_{\mu}$
decay from CLEO \cite{Cleo} ( See also Ref. \cite{With} ), we obtain the
following theoretical rates
\beqa
{\cal B}(B^o \ra D^- D^+_s) = 1.22 \hs{2mm} |a_1 - P|^2 \hs{2mm} 10^{-2} \no \\
{\cal B}(B^o \ra D^- {D^*_s}^+) = 0.87 \hs{2mm} |a_1 - P|^2 \hs{2mm} 10^{-2}
\no \\
{\cal B}(B^o \ra {D^*}^- D^+_s) = 0.83 \hs{2mm} |a_1 - P|^2 \hs{2mm} 10^{-2}
\label{eq12} \\
{\cal B}(B^o \ra {D^*}^- {D^*_s}^+) = 2.13 \hs{2mm} |a_1 - P|^2 \hs{2mm}
10^{-2} \no
\eeqa
The penguin corrections, P, are small for all modes but $D^- D^+_s$ where we
have the estimate
$| a_1 - P |^2 \simeq ( a_1 - 0.15 )^2 $.

The only available experimental data are from ARGUS (See Ref.\cite{Browder})
with a typical number of events between 2 and 4 - e.g.  very poor statistics .
The one standard deviation intersection range for the four modes in (12) is,
\beq
0.83 \leq a_1 \leq 1.12 \label{eq13}
\eeq
which confirms a value of $a_1$ close to unity.

Neglecting the ${D}^- - \ol{D}^o$ and ${D^-}^* - {\ol{D}^o}^*$ mass differences
we can use the same theoretical expression in the analysis of the decays
$B^+_u \ra \ol{D}^o ({\ol{D}^*}^o) + D^+_s({D^*_s}^+) $.
Again the only available data are from ARGUS  (See Ref.\cite{Browder}) and the
one standard intersection range for the four modes is

\beq
1.03 \leq a_1 \leq 1.22 \label{eq14}.
\eeq
Though the uncertainties in (13) and (14) are large, these values of $a_1$ are
consistent with those obtained from the other Class I processes, $ B^o \ra
\pi^+ (\rho^+) + D^- ( {D^*}^- )$, eq.(4).
This could be interpreted to imply that W-exchange contribution is not large.

\vs{4mm}
We now turn to an independent determination of $|a_2|$ from the only
colour-suppressed mode that has been observed, namely, that involving a
charmonium state and a strange meson :
$B \ra J/\Psi + K (K^*)$.

Best statistics exist \cite{Browder} for $B \ra J/\Psi + K $ in charged and
neutral modes.
Using our set of parameters (1) we obtain for $B \ra J/\Psi + K $ the following
branching ratio \cite{GKP},
\beq
{\cal B}(B \ra J/\Psi + K) = 2.62 \hs{2mm} a_2^2 \hs{2mm}
|F_1^{BK}(m^2_{J/\Psi})|^2 \hs{2mm} \cdot 10^{-2} \label{eq15}
\eeq
If we use the World average data \cite{Browder}
for the charged and neutral modes, we obtain, with one standard deviation
error,
the following intersection \cite{GKP}
\beq
|a_2| \hs{2mm} F_1^{BK}(m^2_{J/\Psi}) = 0.190 \pm 0.005  \label{eq16}.
\eeq
The resulting values of $|a_2|$ in the three models we have considered are,
\beqa
|a_2| & = &    0.338 \pm 0.009 \hs{20mm}{\rm BSWI} \no \\
& = & 0.228 \pm 0.006 \hs{20mm} {\rm BSWII} \label{eq17} \\
& = & 0.262 \pm 0.007 \hs{20mm} {\rm CDDFGN} \no
\eeqa

However, a note of caution :
In \cite{GKP} we have shown that the $B \ra K^*$ transition form factors
generated by these three models are in conflict with the longitudinal
polarization fraction measured \cite{Browder}.
Thus, if the polarization data are correct, one ought to exercise caution in
trusting estimates of $|a_2|$ from $B \ra J/\Psi + K (K^*)$.

\vs{5mm}
Summarizing, we wish to emphasize the difference between our approach and that
adopted by others in the past \cite{{Neubert},{Stone},{Browder}} who determine
$a_1$ from a global fit to the Class I processes,
$B^o \ra \pi^+ (\rho^+) + D^- ({D^*}^-)$, and $|a_2|$ from the Class II
process, $B \ra J/\Psi + K ( K^* )$.
The size and the sign of $a_2/a_1$ is then determined from a global fit to the
ratios of Class III to Class I rates as defined in
\cite{{Neubert},{Stone},{Browder}}.

We, on the other hand, do not attempt a global fit.
We determine $a_1$ from the Class I processes, $B^o \ra \pi^+ (\rho^+) + D^-
({D^*}^-)$, by finding a common intersection.
And, independently, we also find an allowed domain $( a_1, a_2 )$ from the
Class III processes,  $B^{+} \ra \pi^{+} (\rho^+) + \ol{D}^o ( \ol{D}^{{o}^*}
)$. We find that to one standard deviation there is no common intersection of
the allowed band of $a_1$ from Class I decays and the allowed $( a_1, a_2 )$
domain from Class III decays. Only at two standard deviation level  do we find
such a common intersection.
The solutions quoted in \cite{Browder} lie in this, rather limited, region in
$( a_1, a_2 )$ plane.

The reason we do not use $B \ra J/\Psi + K ( K^* )$ to extract $|a_2|$ has been
stated in \cite{GKP}
where it has been shown that the longitudinal polarization fraction,
$\Gamma_{L}/\Gamma$,
measured at ARGUS and CLEO \cite{Browder},
does not agree with the prediction of any of the six models  considered there,
which include the three models  considered here.
Thus evaluation of $|a_2|$ using form factors from models that do not correctly
reproduce $\Gamma_{L}/\Gamma$ has to be treated with caution.

We suggest that the following three measurements be done :
First, as explained in the text, one needs an independent measurement of the
branching ratio
 for $B^+ \ra \rho^+ \ol{D}^o$ which was crucial in determining the boundaries
of the allowed $( a_1, a_2 )$ domain from Class III processes.
Second, a high-statistics measurement of the branching ratios of the Class I
process,
$B \ra \ol{D} (\ol{D}^*) + D_s (D_s^*)$, in both charged states would yield an
independent determination of $a_1$.
This process is free of W-annihilation contamination, though there is a small,
but controlable, penguin contribution.
Third, a measurement of the longitudinal polarization fraction for the Class
III process, $B^{+} \ra  \rho^+{\ol{D}^*}^o$, would be very welcome as it would
be sensitive to the size and the sign of $a_2/a_1$.

%
%
%
%

\vspace{1cm}
\hspace{1cm} \Large{} {\bf Acknowledgements}    \vspace{0.5cm}

\normalsize{
A. N. K wishes to thank the Laboratoire de Physique Th\'eorique et Hautes
Energies
in Paris for their hospitality and the Natural Sciences and Engineering
Research Council of Canada
for a research grant which partly supported this research.

Y. Y. K would like to thank the Commissariat a l'Energie Atomique of France for
award of a fellowship
and especially G. Cohen-Tannoudji.
}

\newpage
%

\newpage
\section*{Figure captions}
\normalsize
\vspace{0.5cm}

{\bf Figure 1.} Allowed ranges for $a_1$ and $a_2$ with one standard deviation
experimental errors.  The two horizontal lines are from HQET II, Eq.(4).
The allowed domain, shown shaded, is from the solution to Eq.(7) and (8) with
one standard deviation errors where we have used BSWI model for $B \ra \pi
(\rho)$ transitions. \\

{\bf Figure 2.} Same as Fig. 1 with BSWII model for $B \ra \pi (\rho)$
transitions. \\

{\bf Figure 3.} Same as Fig. 1 with CDDFGN model for $B \ra \pi (\rho)$
transitions. \\

{\bf Figure 4.} Allowed ranges for $a_1$ and $a_2$ with two standard deviations
experimental errors.  The two horizontal lines are from HQET II, Eq.(4).
The allowed domain, shown shaded, is from the solution to Eq.(7) and (8) with
two standard deviations errors where we have used BSWII model for $B \ra \pi
(\rho)$ transitions.
The intersection of the two is the deeper shaded wedge.

\end{document}